\documentclass[lettersize,journal]{IEEEtran}
\usepackage{amsmath,amsfonts}
\usepackage{amssymb}
\usepackage{algorithmic}
\usepackage{algorithm}
\usepackage{array}
\usepackage[caption=false,font=normalsize,labelfont=sf,textfont=sf]{subfig}
\usepackage{textcomp}
\usepackage{stfloats}
\usepackage{url}
\usepackage{verbatim}
\usepackage{graphicx}
\usepackage{cite}
\newtheorem{remark}{Remark}
\usepackage{booktabs}
\usepackage{multirow}
\usepackage{arydshln}
\usepackage{makecell}
\begin{document}

\title{Massive MIMO CSI Feedback with Spiking Neural Networks}

\author{
	Yanzhen Liu
	and Geoffrey Ye Li,~\IEEEmembership{Fellow,~IEEE}%
	\thanks{Y. Liu and G. Y. Li are with the Department
		of Electrical and Electronic Engineering, Imperial College London,
		London, UK, (e-mail: yanzhen.liu22@ic.ac.uk, geoffrey.li@ic.ac.uk).}
}

\maketitle

\begin{abstract}
	Deep learning-based channel state information (CSI) feedback has achieved empirical success in massive multiple-input multiple-output (MIMO) systems. However, existing approaches largely rely on dense artificial neural networks (ANNs), whose computational overhead limits their practical applications. In this article, we exploit bio-inspired spiking neural networks (SNNs) for massive MIMO CSI feedback, referred to as SpikingCSINet, where both the feedback and the main network computations are implemented through spikes. To overcome the information bottleneck of binary spikes in high-dimensional reconstruction, we develop a progressive residual (PR) architecture that exploits the natural temporal dimension of SNNs, encoding successive residuals across time steps to enhance information compactness. Experiments on the COST 2100 benchmark show that SpikingCSINet attains a more favorable performance-efficiency tradeoff than lightweight convolutional baselines. Moreover, it achieves performance competitive with Transformer-based feedback while reducing energy consumption by over $93\%$.
\end{abstract}

\section{Introduction}
Massive multiple-input multiple-output (MIMO) is a critical technique in modern wireless communication systems. With downlink channel state information (CSI), the base station (BS) can perform precoding to suppress inter-user interference and enhance system throughput \cite{larsson2014massive}. As a result, the performance of massive MIMO systems relies heavily on accurate CSI acquisition at the BS.

In time-division duplex (TDD) systems, the BS can directly estimate the downlink CSI from uplink pilots due to reciprocity. In contrast, in frequency-division duplex (FDD) systems where reciprocity does not hold, the CSI is estimated at the user terminal (UT) side and then fed back to the BS. Due to the large dimensionality of CSI in massive MIMO systems, directly feeding it back would consume substantial uplink bandwidth, making CSI compression indispensable \cite{rao2014distributed}.

To this end, a number of solutions have been proposed for massive MIMO CSI feedback. Early works mainly exploited the channel structure and developed compressed sensing-based approaches \cite{rao2014distributed}. However, these methods usually require heavy iterative reconstruction~\cite{guo2025deep}. To overcome their limitations, deep learning-based CSI compression methods built on artificial neural networks (ANNs) have been developed, which compress the CSI more effectively by learning representations directly from data and have become an important research direction~\cite{csinet,crnet,clnet,csinetplus,transnet}. Works based on convolutional architectures such as CsiNet~\cite{csinet}, CsiNet+~\cite{csinetplus}, CRNet~\cite{crnet}, and CLNet~\cite{clnet}  demonstrated the superiority of deep learning-based compression over compressed sensing-based methods. The Transformer-based TransNet~\cite{transnet} adopts a two-layer multi-head architecture and outperforms previous convolutional baselines.

Despite these advances, existing CSI feedback methods still rely heavily on dense floating-point operations, whose computational costs pose a limitation under energy-constrained conditions. Spiking neural networks (SNNs), a class of bio-inspired neural networks that process information through discrete spikes~\cite{roy2019towards}, provide a promising alternative due to their high energy efficiency. However, for SNNs, reconstruction-oriented tasks are particularly challenging and underexplored \cite{ouyang2026spikingir} because the binary nature of spikes imposes an inherent information bottleneck. Furthermore, massive MIMO CSI feedback is essentially a high-dimensional continuous-valued reconstruction problem under stringent bandwidth budgets.

In this article, we develop a new SNN-based architecture for CSI feedback, named SpikingCSINet. This framework differs radically from the existing ANN-based approaches by representing CSI as dynamic spike trains with spatio-temporal correlation rather than static, quantized floating-point codewords. To mitigate the reconstruction performance loss caused by spike-based representation, we develop a progressive residual (PR) feedback scheme, which successively encodes the residual channel matrix across time steps and substantially improves the information compactness of spike-based feedback. Simulation results demonstrate that the proposed SNN-based solution outperforms lightweight convolutional architectures while achieving performance comparable to that of Transformer-based models. In terms of energy consumption, SpikingCSINet is significantly more efficient than Transformer-based methods and is particularly advantageous over lightweight ANN-based approaches under low feedback-bit budgets. The main contributions of this work include presenting the first attempt at SNN-based CSI feedback and developing a PR scheme to enable effective spike-based reconstruction.

\section{System Model}

We consider an FDD massive MIMO CSI feedback system, where the UT and the BS are equipped with a single antenna and $N_t$ antennas, respectively. An orthogonal frequency-division multiplexing (OFDM) system with $N_c$ subcarriers is assumed. In the downlink, the BS transmits pilot signals to the UT, and the UT estimates the channel matrix
\begin{equation}
	\tilde{\mathbf{H}}_c \in \mathbb{C}^{N_c \times N_t},
\end{equation}
which needs to be fed back from the UT to the BS. To reduce the uplink overhead of the high-dimensional channel matrix, the channel is first transformed into the angle-delay domain as
\begin{equation}
	\mathbf{H}_c = \mathbf{F}_d \tilde{\mathbf{H}}_c \mathbf{F}_a^H,
\end{equation}
where $\mathbf{F}_d \in \mathbb{C}^{N_c \times N_c}$ and $\mathbf{F}_a \in \mathbb{C}^{N_t \times N_t}$ are discrete Fourier transform (DFT) matrices. Owing to the sparsity of massive MIMO channels in the angle-delay domain, only the first $N_s$ rows are retained \cite{csinet}, yielding the truncated channel matrix
\begin{equation}
	\mathbf{H} \in \mathbb{C}^{N_s \times N_t}.
\end{equation}

The truncated channel matrix still has a relatively large dimension and needs to be further compressed. In this work, we adopt a deep learning-based CSI compression framework. Let $f_{\mathrm{enc}}(\cdot; \boldsymbol{\theta}_{\mathrm{E}})$ and $f_{\mathrm{dec}}(\cdot; \boldsymbol{\theta}_{\mathrm{D}})$ denote the encoder and decoder, respectively, where $\boldsymbol{\theta}_{\mathrm{E}}$ and $\boldsymbol{\theta}_{\mathrm{D}}$ are the learnable parameters of the encoder and decoder, respectively. The UT compresses the channel into a low-dimensional codeword
\begin{equation}
	\mathbf{c} = f_{\mathrm{enc}}(\mathbf{H}; \boldsymbol{\theta}_{\mathrm{E}}),
\end{equation}
which is then fed back to the BS. Subsequently, the BS reconstructs the channel as
\begin{equation}
	\hat{\mathbf{H}} = f_{\mathrm{dec}}(\mathbf{c}; \boldsymbol{\theta}_{\mathrm{D}}).
\end{equation}	

The model is trained by minimizing the mean-squared error (MSE) between the reconstructed channel $\hat{\mathbf{H}}$ and the ground truth $\mathbf{H}$:
\begin{equation}
	\min_{\boldsymbol{\theta}_{\mathrm{E}},\,\boldsymbol{\theta}_{\mathrm{D}}}
	\left\| \mathbf{H} - f_{\mathrm{dec}}(f_{\mathrm{enc}}(\mathbf{H}; \boldsymbol{\theta}_{\mathrm{E}}); \boldsymbol{\theta}_{\mathrm{D}}) \right\|_F^2.
\end{equation}

\section{SpikingCSINet}
In this section, we introduce the proposed SNN-based CSI feedback scheme, SpikingCSINet. We first present the spiking neuron model and then describe the base architecture and the PR feedback scheme. The training method and the energy consumption model are also presented in this section.
\subsection{Spiking Neuron Model}
In this work, we adopt the leaky integrate-and-fire (LIF) model \cite{neftci2019surrogate}. The dynamics of an LIF neuron can be written as
\begin{align}
	v[t^-] &= \left(1 - \frac{1}{\tau}\right) v[t-1] + i[t], \label{v_charge} \\
	s[t]   &= \Theta\!\left(v[t^-] - v_{\mathrm{th}}\right), \label{s_gen} \\
	v[t]   &= v[t^-]\left(1 - s[t]\right) + v_{\mathrm{reset}}\, s[t]. \label{v_reset}
\end{align}
Equation~\eqref{v_charge} describes the charging process, where $v[t^-]$ denotes the pre-reset membrane potential at time step $t$, and $v[t-1]$ denotes the post-reset membrane potential at time step $t-1$. Variable $i[t]$ is the input current, and $\tau > 1$ is the membrane time constant that controls the leakage rate. Equation~\eqref{s_gen} defines the spike generation process, where $\Theta(\cdot)$ denotes the Heaviside step function and $v_{\mathrm{th}}$ is the firing threshold. Equation~\eqref{v_reset} describes the reset process: when a neuron fires, its membrane potential is reset to $v_{\mathrm{reset}}$. By default, we set $v_{\mathrm{th}} =1$ and $v_{\mathrm{reset}}=0$.

The input current is computed as a weighted sum of the spikes from the preceding layer. Specifically, consider the $l$-th fully connected layer, and let $\mathbf{i}^l[t] \in \mathbb{R}^{N_l}$ denote its input current. It is given by
\begin{equation}
	\mathbf{i}^l[t] = \mathbf{W}^{l}\, \mathbf{s}^{l-1}[t],
\end{equation}
where $\mathbf{W}^{l} \in \mathbb{R}^{N_{l} \times N_{l-1}}$ denotes the synaptic weights of layer $l$, and $\mathbf{s}^{l-1}[t] \in \{0,1\}^{N_{l-1}}$ denotes the spikes generated by the neurons in layer $l-1$ at time step $t$. Notably, since the elements of $\mathbf{s}^{l-1}[t]$ are binary, the matrix-vector multiplication reduces to summing the weights associated with non-zero spikes, which substantially lowers the energy consumption compared with the floating-point multiply-accumulate (MAC) operations in ANNs.

\subsection{Network Architecture}
\begin{figure*}[t]
	\centering
	\includegraphics[width=1.0\linewidth]{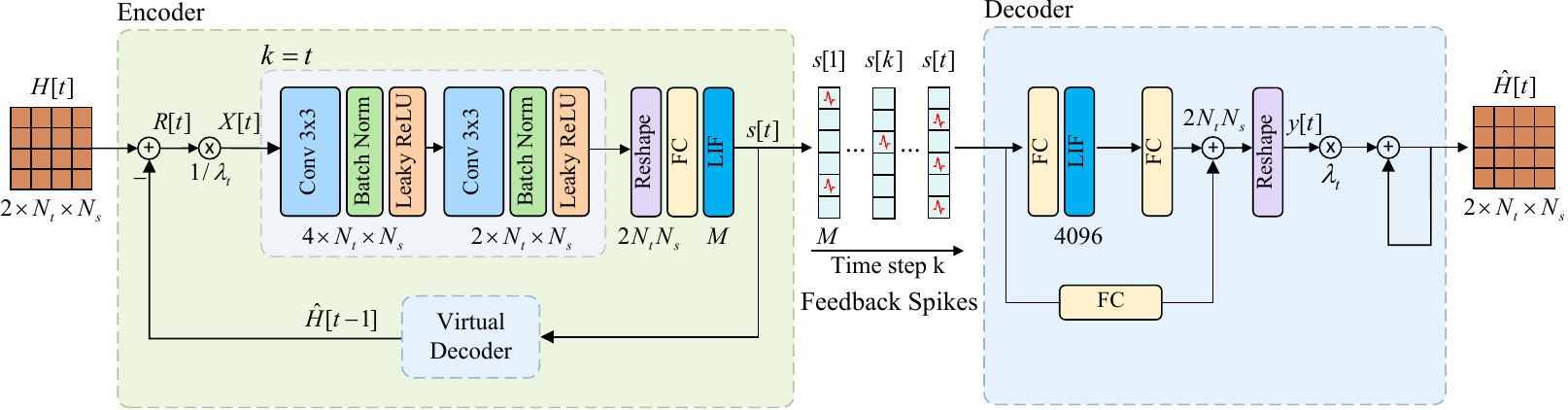}
	\caption{Architecture of the proposed SpikingCSINet with progressive residual (PR) feedback.}
	\label{fig:network_architecture}
	\vspace{-1em}
\end{figure*}

We illustrate the architecture of the proposed SpikingCSINet in Fig.~\ref{fig:network_architecture}, which is described in detail below.

\subsubsection{Base Architecture}

We adopt simple yet effective components to build the SNN for CSI feedback. 
The encoder follows the classical CsiNet structure \cite{csinet} and consists of two convolutional layers, each with a $3\times 3$ kernel, stride 1, and padding 1. 
Each convolutional layer is followed by batch normalization and a Leaky ReLU activation with a slope of 0.3. 
Compared with the original CsiNet encoder, the main difference is that the first convolutional layer uses four output channels instead of two to improve the representation capacity of the encoder. The extracted features are then passed through a fully connected (FC) layer followed by LIF neurons, and the resulting output spikes are transmitted to the BS for CSI reconstruction. 

At the BS, the received spikes are fed into the decoder to reconstruct the CSI matrix. 
The backbone of the decoder consists of a single hidden FC layer followed by LIF neurons. 
The spikes produced by this hidden layer are then passed through an output FC layer to yield the backbone reconstruction. With binary spike inputs, this output FC layer functions as a spike-indexed dictionary, where active spikes select the corresponding columns of the weight matrix and sum them to reconstruct the CSI. 
Motivated by this interpretation, we set the dimension of the decoder LIF neurons to 4,096 to enhance the representation capacity of the network. 
Despite this large dimension, the event-driven nature of spiking computation keeps the computational cost low, as will be shown in Section~\ref{sec:simulation}. 
In addition, we introduce a skip residual connection that directly maps the received spikes to the reconstruction output, which helps preserve the originally fed-back information. 
The outputs of the backbone branch and the skip branch are summed to yield the final reconstruction.

\subsubsection{Progressive Residual Refinement}
Unlike ANNs, SNNs naturally operate in the spatio-temporal domain. Therefore, the input at each time step must be carefully designed.
Classical approaches convert the input into spikes through schemes such as Poisson coding or latency coding~\cite{auge2021survey,sun2024delay}, or directly feed a static analog input through an input spiking layer that generates spikes across time steps \cite{wu2019direct}. These methods do not provide new information across time steps, and thus suffer from the spike-induced information bottleneck and are not well suited to the CSI reconstruction task.

In this work, we instead formulate CSI feedback as a progressive refinement process, which is fundamentally different from existing ANN-based approaches that treat it as a single-shot reconstruction problem.
Specifically, as illustrated in Fig.~\ref{fig:network_architecture}, we deploy a virtual decoder at the UT, which is identical to the one at the BS.
Let $\bar{\mathbf{H}}[t-1]$ denote the reconstructed channel matrix obtained up to time step $t-1$. At time step $t$, we compute the residual between the ground-truth channel, $\mathbf{H}$, and the current reconstruction, $\bar{\mathbf{H}}[t-1]$,
\begin{equation}
	\mathbf{R}[t] = \mathbf{H} - \bar{\mathbf{H}}[t-1].
\end{equation}
Since the residual magnitude tends to decrease over time, we scale it to bring the encoder input back to a comparable numerical range:
\begin{equation}
	\mathbf{X}[t] = \frac{\mathbf{R}[t]}{\lambda[t]},
\end{equation}
where $\lambda[t]$ is a scaling factor computed as
\begin{equation}
	\lambda[t] =
	\left(
	\frac{
		\sum_{\mathbf{H} \in \bar{\mathcal{H}}}
		\|\mathbf{R}[t]\|_F^2
	}{
		\sum_{\mathbf{H} \in \bar{\mathcal{H}}}
		\|\mathbf{H}\|_F^2
	}
	\right)^{1/2},
\end{equation}
with $\bar{\mathcal{H}} \subseteq \mathcal{H}$ being a subset of channel samples (e.g., taken from several mini-batches).

The scaled input is then fed into the encoder to produce the transmitted spikes:
\begin{equation}
	\mathbf{s}[t] =
	f_{\mathrm{enc}}\!\left(
	\mathbf{X}[t];
	\boldsymbol{\theta}_{\mathrm{E}},
	\boldsymbol{\omega}_{\mathrm{E}}[t]
	\right) \in \{0,1\}^{M},
\end{equation}
where $\boldsymbol{\omega}_{\mathrm{E}}[t]$ denotes the encoder state variables, such as membrane potentials, and $M$ denotes the compressed codeword size. 
Here, we use a separate set of convolutional encoder parameters at each time step because the residual inputs exhibit different characteristics across time steps. Specifically, we define $\boldsymbol{\theta}_{\mathrm{E}} \triangleq \{\{\boldsymbol{\theta}_{\mathrm{E}}^{\mathrm{conv}}[t]\}_{t=1}^{T},\boldsymbol{\theta}_{\mathrm{E}}^{\mathrm{fc}}\}$ with $\boldsymbol{\theta}_{\mathrm{E}}^{\mathrm{conv}}[t]$ denoting the encoder parameters of convolutional layers at time step $t$, and $\boldsymbol{\theta}_{\mathrm{E}}^{\mathrm{fc}}$ denoting the encoder parameters of the FC layer.
As the convolutional kernels are lightweight, this design introduces only negligible overhead.

Upon receiving the spikes, the decoder processes them to produce the step-wise reconstruction output:
\begin{equation}
	\mathbf{y}[t] =
	f_{\mathrm{dec}}\!\left(
	\mathbf{s}[t];
	\boldsymbol{\theta}_{\mathrm{D}},
	\boldsymbol{\omega}_{\mathrm{D}}[t]
	\right),
\end{equation}
where $\boldsymbol{\omega}_{\mathrm{D}}[t]$ denotes the decoder state variables.
The estimated channel is then progressively refined by accumulating the rescaled outputs:
\begin{equation}
	\bar{\mathbf{H}}[t]
	=
	\sum_{k=1}^{t}
	\lambda[k]\, \mathbf{y}[k],
\end{equation}
with $\bar{\mathbf{H}}[0]=\mathbf{0}$ as the initial condition and $\hat{\mathbf{H}} = \bar{\mathbf{H}}[T]$ as the final output.

\begin{remark}
This PR architecture is intrinsically tailored to SNN-based feedback. 
The temporal unfolding of SNNs naturally supports progressive residual encoding. The generated spikes serve as the minimal feedback units and avoid additional quantization. 
Moreover, the refinement is driven by sparse spike events rather than dense floating-point operations, making it energy-efficient even across multiple time steps. 
In this way, PR converts the temporal dynamics of SNNs into an information-refinement mechanism that mitigates the spike-induced bottleneck while keeping bandwidth usage and energy consumption low.
\end{remark}

\vspace{-0.5em}
\subsection{Training Strategy}
We train SpikingCSINet using backpropagation through time (BPTT), where the non-differentiable spiking function is approximated by the surrogate gradient method~\cite{neftci2019surrogate}, with the arctangent function as the default surrogate. Defining $\boldsymbol{\theta} \triangleq \{\boldsymbol{\theta}_{\mathrm{E}},\boldsymbol{\theta}_{\mathrm{D}}\}$ as the collection of learnable parameters,
the training objective is defined as
\begin{equation}
	L(\boldsymbol{\theta}) = \|\bar{\mathbf{H}}[T] - \mathbf{H}\|_F^2 
	+ \alpha \sum_{t=1}^{T-1} \|\bar{\mathbf{H}}[t] - \mathbf{H}\|_F^2.
\end{equation}
The first term measures the final reconstruction error, while the second term penalizes the intermediate reconstruction errors, with $\alpha \geq 0$ being a weighting coefficient that controls their relative contribution.

\vspace{-0.5em}
\subsection{Energy Consumption}
Due to the event-driven nature of spike-based computation, the number of floating-point operations (FLOPs) is not a suitable metric for evaluating SNN models.
Instead, their computational cost is generally measured by the total number of accumulation (AC) operations and MAC operations~\cite{deng2020rethinking}.
The total energy consumption is computed as
\begin{equation}
	E = E_{\mathrm{mac}} N_{\mathrm{mac}} + E_{\mathrm{ac}} N_{\mathrm{ac}},
\end{equation}
where $E_{\mathrm{mac}}$ and $E_{\mathrm{ac}}$ denote the energy costs per MAC and per AC operation, respectively.
The encoder of SpikingCSINet uses floating-point MAC operations in its convolutional layers and final FC layer.
The decoder consists of spike-based FC layers, in which the computation reduces to AC operations. For each spike-based FC layer, the number of AC operations is given by
\begin{equation}
	N_{\mathrm{ac}}^{\mathrm{s},\mathrm{fc}}
	=
	\rho M_{\mathrm{in}} M_{\mathrm{out}},
\end{equation}
where $\rho$ denotes the average firing rate of the input spikes, and $M_{\mathrm{in}}$ and $M_{\mathrm{out}}$ denote the input and output dimensions of the FC layer.

\section{Simulation Results}
\label{sec:simulation}
\begin{table*}[t]
	\centering
	\caption{NMSE (dB) and energy consumption under different compression ratios (CRs) and feedback bits ($T/b$) on the COST 2100 dataset.}
	\label{tab:all_nmse_energy_compact}
	\vspace{-1em}
	\small
	\setlength{\tabcolsep}{3pt}
	\renewcommand{\arraystretch}{0.95}
	\resizebox{\textwidth}{!}{
		\begin{tabular}{llccccccccccccccccc}
			\toprule
			\multirow{2}{*}{Methods} & \multirow{2}{*}{$T/b$}
			& \multicolumn{3}{c}{$\text{CR}=4$}
			& \multicolumn{3}{c}{$\text{CR}=8$}
			& \multicolumn{3}{c}{$\text{CR}=16$}
			& \multicolumn{3}{c}{$\text{CR}=32$}
			& \multicolumn{3}{c}{$\text{CR}=64$} \\
			\cmidrule(lr){3-5}\cmidrule(lr){6-8}\cmidrule(lr){9-11}\cmidrule(lr){12-14}\cmidrule(lr){15-17}
			& & \multicolumn{2}{c}{NMSE} & \multirow{2}{*}{Energy}
			& \multicolumn{2}{c}{NMSE} & \multirow{2}{*}{Energy}
			& \multicolumn{2}{c}{NMSE} & \multirow{2}{*}{Energy}
			& \multicolumn{2}{c}{NMSE} & \multirow{2}{*}{Energy}
			& \multicolumn{2}{c}{NMSE} & \multirow{2}{*}{Energy} \\
			\cmidrule(lr){3-4}\cmidrule(lr){6-7}\cmidrule(lr){9-10}\cmidrule(lr){12-13}\cmidrule(lr){15-16}
			& & Indoor & Outdoor &
			& Indoor & Outdoor &
			& Indoor & Outdoor &
			& Indoor & Outdoor &
			& Indoor & Outdoor & \\
			\midrule
			
			\multirow{3}{*}{\makecell[l]{SpikingCSINet\\(proposed)}}
			& 2 & \textbf{-17.10} & \textbf{-8.22} & 8.05\,$\mu$J & \textbf{-13.60} & \textbf{-6.29} & 4.52\,$\mu$J & \textbf{-9.95} & \textbf{-4.50} & 2.76\,$\mu$J & \textbf{-7.11} & \textbf{-3.05} & 1.89\,$\mu$J & \textbf{-3.97} & \textbf{-1.68} & 1.44\,$\mu$J \\
			& 4 & \textbf{-24.04} & \textbf{-10.07} & 16.11\,$\mu$J & \textbf{-17.85} & \textbf{-7.77} & 9.02\,$\mu$J & \textbf{-12.34} & \textbf{-5.31} & 5.53\,$\mu$J & \textbf{-8.47} & \textbf{-3.36} & 3.76\,$\mu$J & \textbf{-5.08} & \textbf{-1.95} & 2.87\,$\mu$J \\
			& 6 & \textbf{-27.83} & \textbf{-11.56} & 24.20\,$\mu$J & \textbf{-19.75} & \textbf{-8.75} & 13.52\,$\mu$J & \textbf{-13.55} & \textbf{-5.77} & 8.27\,$\mu$J & \textbf{-9.90} & \textbf{-3.52} & 5.63\,$\mu$J & \textbf{-6.45} & \textbf{-2.08} & 4.30\,$\mu$J \\
			\midrule
			
			\multirow{3}{*}{SpikingCSINet-L}
			& 2 & \textbf{-18.17} & \textbf{-10.87} & 8.25\,$\mu$J & \textbf{-14.34} & \textbf{-8.39} & 4.63\,$\mu$J & \textbf{-10.39} & \textbf{-6.07} & 2.83\,$\mu$J & \textbf{-7.27} & \textbf{-4.15} & 1.95\,$\mu$J & \textbf{-3.96} & \textbf{-2.30} & 1.46\,$\mu$J \\
			& 4 & \textbf{-25.51} & \textbf{-12.29} & 16.54\,$\mu$J & \textbf{-18.87} & \textbf{-10.03} & 9.28\,$\mu$J & \textbf{-12.98} & \textbf{-6.64} & 5.71\,$\mu$J & \textbf{-8.61} & \textbf{-4.31} & 3.88\,$\mu$J & \textbf{-5.54} & \textbf{-2.47} & 2.92\,$\mu$J \\
			& 6 & \textbf{-29.71} & \textbf{-13.66} & 24.87\,$\mu$J & \textbf{-20.96} & \textbf{-10.96} & 13.92\,$\mu$J & \textbf{-14.30} & \textbf{-7.26} & 8.50\,$\mu$J & \textbf{-10.24} & \textbf{-4.53} & 5.76\,$\mu$J & \textbf{-6.47} & \textbf{-2.73} & 4.37\,$\mu$J \\
			
			\midrule
			\midrule
			
			\multirow{4}{*}{CsiNet}
			& 2 & -12.44 & -7.26 & 17.33\,$\mu$J & -7.39 & -5.85 & 13.97\,$\mu$J & -5.94 & -3.46 & 12.29\,$\mu$J & -4.43 & -2.10 & 11.46\,$\mu$J & -4.61 & -1.48 & 11.04\,$\mu$J \\
			& 4 & -15.59 & -8.55 & 17.33\,$\mu$J & -10.60 & -7.03 & 13.97\,$\mu$J & -7.77 & -4.20 & 12.29\,$\mu$J & -5.70 & -2.63 & 11.46\,$\mu$J & -5.57 & -1.82 & 11.04\,$\mu$J \\
			& 6 & -17.33 & -8.87 & 17.33\,$\mu$J & -12.04 & -7.43 & 13.97\,$\mu$J & -8.40 & -4.42 & 12.29\,$\mu$J & -6.09 & -2.77 & 11.46\,$\mu$J & -5.80 & -1.91 & 11.04\,$\mu$J \\
			& $\infty$ & -17.36 & -8.75 & 17.33\,$\mu$J & -12.70 & -7.61 & 13.97\,$\mu$J & -8.65 & -4.51 & 12.29\,$\mu$J & -6.24 & -2.81 & 11.46\,$\mu$J & -5.84 & -1.93 & 11.04\,$\mu$J \\
			\midrule
			
			\multirow{4}{*}{CLNet}
			& 2 & -14.12 & -9.79 & 12.88\,$\mu$J & -10.37 & -6.67 & 9.53\,$\mu$J & -8.60 & -4.46 & 7.85\,$\mu$J & -7.20 & -2.88 & 7.01\,$\mu$J & -4.70 & -1.71 & 6.59\,$\mu$J \\
			& 4 & -19.74 & -12.11 & 12.88\,$\mu$J & -13.78 & -7.96 & 9.53\,$\mu$J & -10.46 & -5.33 & 7.85\,$\mu$J & -8.57 & -3.37 & 7.01\,$\mu$J & -6.01 & -2.08 & 6.59\,$\mu$J \\
			& 6 & -24.71 & -12.77 & 12.88\,$\mu$J & -15.19 & -8.24 & 9.53\,$\mu$J & -11.02 & -5.53 & 7.85\,$\mu$J & -8.88 & -3.47 & 7.01\,$\mu$J & -6.30 & -2.18 & 6.59\,$\mu$J \\
			& $\infty$ & -29.16 & -12.88 & 12.88\,$\mu$J & -15.60 & -8.29 & 9.53\,$\mu$J & -11.15 & -5.56 & 7.85\,$\mu$J & -8.95 & -3.49 & 7.01\,$\mu$J & -6.34 & -2.19 & 6.59\,$\mu$J \\
			\midrule
			
			\multirow{4}{*}{CRNet}
			& 2 & -16.82 & -10.45 & 16.39\,$\mu$J & -11.03 & -6.19 & 13.04\,$\mu$J & -8.74 & -4.38 & 11.36\,$\mu$J & -6.93 & -2.71 & 10.52\,$\mu$J & -5.03 & -1.67 & 10.10\,$\mu$J \\
			& 4 & -21.65 & -12.20 & 16.39\,$\mu$J & -14.15 & -7.58 & 13.04\,$\mu$J & -10.79 & -5.22 & 11.36\,$\mu$J & -8.42 & -3.32 & 10.52\,$\mu$J & -6.22 & -2.09 & 10.10\,$\mu$J \\
			& 6 & -25.09 & -12.64 & 16.39\,$\mu$J & -15.57 & -7.96 & 13.04\,$\mu$J & -11.27 & -5.41 & 11.36\,$\mu$J & -8.83 & -3.48 & 10.52\,$\mu$J & -6.46 & -2.19 & 10.10\,$\mu$J \\
			& $\infty$ & -26.99 & -12.70 & 16.39\,$\mu$J & -16.01 & -8.04 & 13.04\,$\mu$J & -11.35 & -5.44 & 11.36\,$\mu$J & -8.93 & -3.51 & 10.52\,$\mu$J & -6.49 & -2.22 & 10.10\,$\mu$J \\
			\midrule
			
			\multirow{4}{*}{CsiNet+}
			& 2 & / & / & / & / & / & / & / & / & / & / & / & / & / & / & / \\
			& 4 & -19.06 & -11.54 & 78.63\,$\mu$J & -15.65 & -7.94 & 75.27\,$\mu$J & -12.58 & -5.49 & 73.60\,$\mu$J & -9.37 & -2.93 & 72.76\,$\mu$J & / & / & / \\
			& 6 & -24.97 & -12.83 & 78.63\,$\mu$J & -18.03 & -8.81 & 75.27\,$\mu$J & -14.02 & -5.75 & 73.60\,$\mu$J & -10.35 & -3.79 & 72.76\,$\mu$J & / & / & / \\
			& $\infty$ & -27.37 & -12.40 & 78.63\,$\mu$J & -18.29 & -8.72 & 75.27\,$\mu$J & -14.14 & -5.73 & 73.60\,$\mu$J & -10.43 & -3.40 & 72.76\,$\mu$J & / & / & / \\
			\midrule
			
			\multirow{4}{*}{TransNet}
			& 2 & -18.29 & -11.26 & 124.15\,$\mu$J & -13.94 & -7.99 & 120.80\,$\mu$J & -10.04 & -6.38 & 119.12\,$\mu$J & -7.32 & -2.52 & 118.28\,$\mu$J & -3.34 & -1.64 & 117.86\,$\mu$J \\
			& 4 & -24.48 & -13.93 & 124.15\,$\mu$J & -18.93 & -9.58 & 120.80\,$\mu$J & -13.17 & -7.50 & 119.12\,$\mu$J & -8.91 & -3.59 & 118.28\,$\mu$J & -5.34 & -2.31 & 117.86\,$\mu$J \\
			& 6 & -29.36 & -14.85 & 124.15\,$\mu$J & -21.95 & -10.00 & 120.80\,$\mu$J & -14.56 & -7.82 & 119.12\,$\mu$J & -10.01 & -4.06 & 118.28\,$\mu$J & -5.99 & -2.61 & 117.86\,$\mu$J \\
			& $\infty$ & -32.38 & -14.86 & 124.15\,$\mu$J & -22.91 & -9.99 & 120.80\,$\mu$J & -15.00 & -7.82 & 119.12\,$\mu$J & -10.49 & -4.13 & 118.28\,$\mu$J & -6.08 & -2.62 & 117.86\,$\mu$J \\
			\bottomrule
		\end{tabular}
	}
		\vspace{1.2mm}
\begin{minipage}{\textwidth}
	\scriptsize
	1) $T$ and $b$ denote the SNN time steps and the ANN quantization bit-width, respectively, both reflecting the feedback bit budget. $\infty$ denotes the full-precision ANN-based schemes. The compression ratio is defined as $\text{CR} \triangleq \frac{2N_tN_s}{M}$.\\
	2) For CsiNet+, we adopt the performance reported in the original paper, and ``/'' indicates that the performance was not reported. The original papers of CsiNet, CRNet, CLNet, and TransNet did not report quantized performance. We train these models using their official code and fine-tune them with the straight-through estimator to obtain quantized results.
\end{minipage}\vspace{-1em}
\end{table*}

In this section, we evaluate the proposed SpikingCSINet through simulation.
We adopt the standard COST 2100 benchmark~\cite{liu2012cost2100} for fair comparison with previous works~\cite{csinet,csinetplus,crnet,clnet,transnet}.
The dataset contains indoor and outdoor scenarios. Each consists of $10{,}000$ training samples, $30{,}000$ validation samples, and $20{,}000$ testing samples.
The truncated channel matrix has dimensions $N_s = N_t = 32$ and the inputs are scaled to the range $[-25, 25]$ to drive sufficient spiking activity.
Moreover, since the spike-based model tends to overfit in the indoor scenario, we adopt a simple data augmentation strategy that multiplies the channel by a random phase-rotation coefficient $\gamma = \exp(-2\pi \mathrm{j} k / K)$, where $k$ is uniformly drawn from $\{0, 1, \ldots, K-1\}$ and $K = 16$.
By default, we use the Adam optimizer with a learning rate of $0.002$ and a cosine learning rate schedule. The model is trained for $1{,}000$ epochs.
The LIF neurons are configured with $\tau = 2$, and the loss weighting coefficient is set to $\alpha = 0.5$.
For energy estimation, we adopt the standard model based on a 45-nm CMOS process~\cite{horowitz2014computing}, where $E_{\mathrm{mac}} = 3.2\,\mathrm{pJ}$ and $E_{\mathrm{ac}} = 0.1\,\mathrm{pJ}$. The code is available at \url{https://github.com/yl5922/SpikingCSINet}.

Table~\ref{tab:all_nmse_energy_compact} summarizes the normalized MSE (NMSE) performance and energy consumption of all evaluated schemes. Compared with the convolutional architectures, SpikingCSINet reduces the linear-domain NMSE by $54.4\%$, $28.4\%$, $21.6\%$, and $17.2\%$ over CsiNet, CLNet, CRNet, and CsiNet+, respectively, on the indoor dataset. Its outdoor performance, however, is less competitive due to the more complex scattering environment. To close this gap, we further introduce SpikingCSINet-L, an extended variant with the decoder LIF dimension enlarged from $4,096$ to $8,192$. SpikingCSINet-L improves the reconstruction quality in both scenarios, achieving average NMSE reductions of $39.8\%$ and $29.0\%$ over the four convolutional baselines on the indoor and outdoor datasets, respectively. Notably, this gain comes at only a $2.5\%$ average increase in energy consumption over SpikingCSINet since the additional computation is purely spike-driven and benefits from the inherent sparsity of AC operations. 

In terms of energy efficiency, SpikingCSINet is more efficient than the convolutional baselines, especially under high compression ratios and limited feedback budgets. For instance, when $T/b=2$, SpikingCSINet and SpikingCSINet-L reduce the energy consumption by $61.2\%$ and $60.3\%$ on average over CLNet, the most energy-efficient convolutional architecture. Compared with the stronger but more expensive TransNet, SpikingCSINet-L attains comparable reconstruction performance, with a $2.0\%$ average improvement indoors and only a $0.9\%$ degradation outdoors, while reducing the energy consumption by $93.7\%$ on average.

\begin{figure}[t]
	\centering
	\begin{minipage}{0.48\columnwidth}
		\centering
		\includegraphics[width=\linewidth]{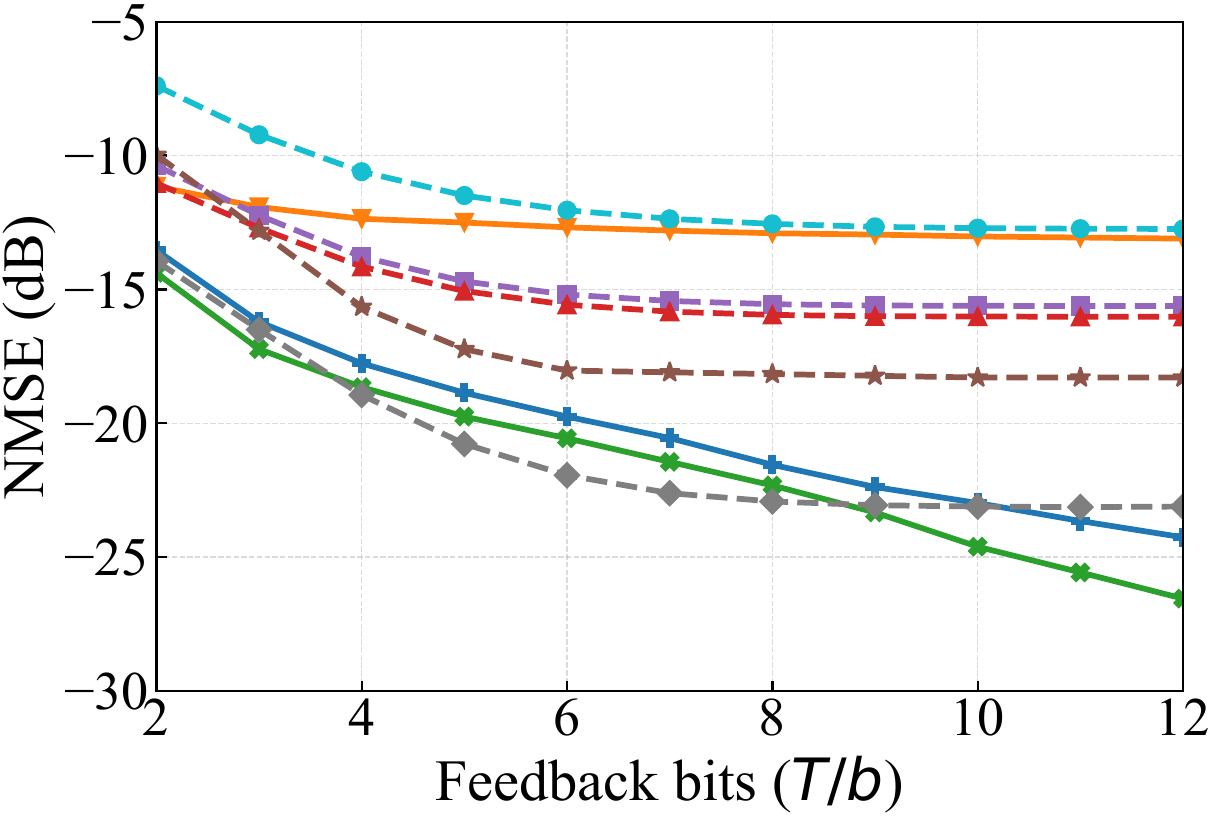}
		\\[-1mm]
		{\footnotesize (a) NMSE}
	\end{minipage}
	\hfill
	\begin{minipage}{0.48\columnwidth}
		\centering
		\includegraphics[width=\linewidth]{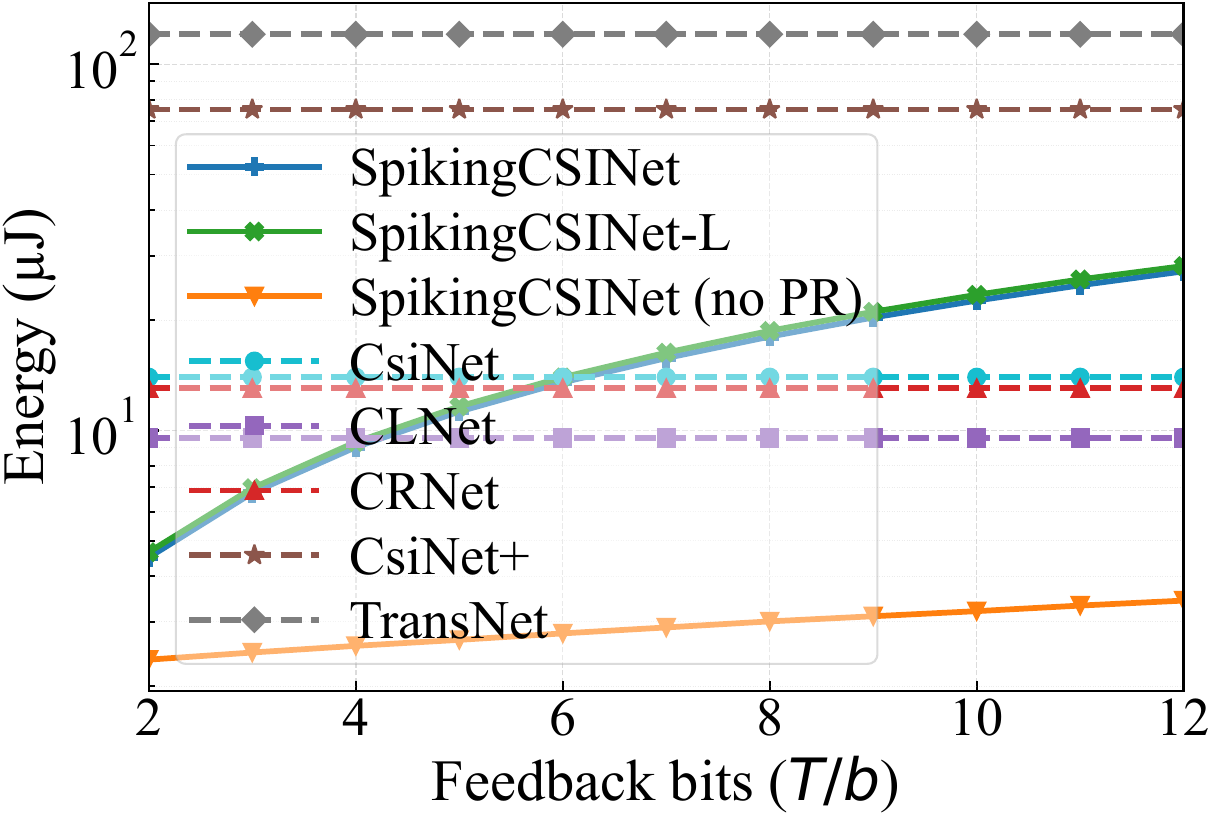}
		\\[-1mm]
		{\footnotesize (b) Energy}
	\end{minipage}
	\caption{(a) NMSE and (b) energy consumption versus feedback bits under $\text{CR}=8$ in the indoor scenario.}
	\label{fig:t_curve}
	\vspace{-1em}
\end{figure}

Fig.~\ref{fig:t_curve} compares the NMSE performance and energy consumption versus different feedback bit budgets ($T/b$). We include SpikingCSINet without the PR architecture as an ablation, denoted as SpikingCSINet (no PR) in the figure.
As shown in Fig.~\ref{fig:t_curve}(a), SpikingCSINet continues to improve as $T/b$ increases, whereas the ANN-based feedback schemes tend to saturate once $T/b$ exceeds $8$. When $T/b$ exceeds $10$, SpikingCSINet surpasses TransNet, and SpikingCSINet-L achieves further improvement. This contrast reflects a fundamental difference between SNN-based and ANN-based quantized feedback: SpikingCSINet distributes the feedback information through successive spatio-temporal coding, whereas ANN-based quantization relies on a static representation whose performance saturates as the bit-width increases. The ablation scheme, SpikingCSINet (no PR), only achieves performance comparable to CsiNet, demonstrating the necessity of PR refinement.
Fig.~\ref{fig:t_curve}(b) further compares the energy consumption. The energy consumption of SpikingCSINet and SpikingCSINet-L increases with the number of time steps due to their successive refinement design, but remains much lower than that of heavier ANN-based schemes, such as TransNet and CsiNet+. When $T/b \leq 4$, they further outperform lightweight convolutional architectures in terms of both effectiveness and efficiency. SpikingCSINet (no PR) achieves the lowest energy consumption, but its limited reconstruction accuracy makes this saving less meaningful.


Table~\ref{tab:param_counts_horizontal_M} reports the model sizes of different schemes under $\text{CR}=8$. Due to its spike-based decoder design, SpikingCSINet uses more parameters than the compared ANN-based schemes. Nevertheless, its parameter size is still within a reasonable range among CSI feedback models. For example, DeepCMC contains approximately $9.95$M parameters, while ConvLSTM-CsiNet contains about $22.30$M parameters under $\text{CR}=8$~\cite{deepcmc,convlstmcsinet}. More importantly, a larger parameter size does not necessarily imply higher energy consumption in SNNs. As shown in Table~\ref{tab:spike_rates_T6}, the average decoder firing rate of SpikingCSINet ranges only from 0.013 to 0.071, and that of SpikingCSINet-L is even lower. Therefore, only a small fraction of synaptic connections are activated at each time step, and most operations can be implemented as low-cost AC operations. In contrast, increasing the parameter size of ANN-based architectures usually leads to a direct increase in dense MAC operations.


\vspace{-1em}
\begin{table}[t]
	\centering
	\caption{Parameter counts of different schemes under $\text{CR}=8$.}
	\label{tab:param_counts_horizontal_M}
	\vspace{-1.5em}
	\huge
	\resizebox{\columnwidth}{!}{%
		\begin{tabular}{lccccccc}
			\toprule
			Scheme & CsiNet & CsiNet+ & CRNet & CLNet & TransNet & SpikingCSINet & SpikingCSINet-L \\
			\midrule
			Param. (M) & 1.05 & 1.07 & 1.05 & 1.05 & 1.32 & 10.49 & 19.93 \\
			\bottomrule
		\end{tabular}%
	}
\end{table}

\begin{table}[t]
	\centering
	\caption{Average decoder LIF-layer firing rates of SpikingCSINet and SpikingCSINet-L under $T=6$.}
	\label{tab:spike_rates_T6}
	\resizebox{\columnwidth}{!}{%
		\begin{tabular}{llccccc}
			\toprule
			Environment & Scheme & $\text{CR}=4$ & $\text{CR}=8$ & $\text{CR}=16$ & $\text{CR}=32$ & $\text{CR}=64$ \\
			\midrule
			\multirow{2}{*}{Indoor}
			& SpikingCSINet   & 0.0710 & 0.0421 & 0.0465 & 0.0463 & 0.0411 \\
			& SpikingCSINet-L & 0.0647 & 0.0343 & 0.0318 & 0.0286 & 0.0252 \\
			\midrule
			\multirow{2}{*}{Outdoor}
			& SpikingCSINet   & 0.0575 & 0.0284 & 0.0202 & 0.0132 & 0.0146 \\
			& SpikingCSINet-L & 0.0222 & 0.0181 & 0.0154 & 0.0096 & 0.0079 \\
			\bottomrule
		\end{tabular}%
	}\vspace{-0.5em}
\end{table}

\section{Conclusion}
In this work, we presented SpikingCSINet, an SNN-based architecture for CSI feedback. A progressive residual architecture is developed to overcome the information bottleneck of binary spikes in high-dimensional reconstruction. Experiments on the COST 2100 dataset demonstrate that, with acceptable memory overhead, SpikingCSINet achieves a more favorable performance-efficiency tradeoff than convolutional architectures while achieving competitive performance compared with Transformer-based architectures at over 93\% lower energy consumption.

\end{document}